# Pressure induced color change and evolution of metallic behavior in nitrogen-doped lutetium hydride

Ying-Jie Zhang, Xue Ming, Qing Li[*], Xiyu Zhu[*], Bo Zheng, Yuecong Liu, Chengping He, Huan Yang, and Hai-Hu Wen[*]

National Laboratory of Solid State Microstructures and Department of Physics, Collaborative Innovation Center of Advanced Microstructures, Nanjing University, Nanjing 210093, China

*Corresponding authors:
liqing1118@nju.edu.cn, zhuxiyu@nju.edu.cn, hhwen@nju.edu.cn

**Abstract:** By applying pressures up to 42 GPa on the nitrogen-doped lutetium hydride ($LuH_{2\pm x}N_y$), we have found a gradual change of color from dark-blue to pink-violet in the pressure region of about 12 GPa to 21 GPa. The temperature dependence of resistivity under pressures up to 50.5 GPa shows progressively optimized metallic behavior with pressure. Interestingly, in the pressure region for the color change, a clear decrease of resistivity is observed with the increase of pressure, which is accompanied by a clear increase of the residual resistivity ratio (*RRR*). Fitting to the low temperature resistivity gives exponents of about 2, suggesting a Fermi liquid behavior in low temperature region. The general behavior in wide temperature region suggests that the electron-phonon scattering is still the dominant one. The magnetoresistance up to 9 tesla in the state under a pressure of 50.5 GPa shows an almost negligible effect, which suggests that the electric conduction in the pink-violet state is dominated by a single band. It is highly desired to have theoretical efforts in understanding the evolution of color and resistivity in this interesting system.

**Introduction**

In 1999, Ginzburg listed about 30 important and interesting problems in physics [1]. The second one on the list is about the room temperature superconductivity. There is no doubt that room temperature superconductivity is one of the dreamful goals for scientists working in the field of superconductivity. However, it seems very difficult to reach this goal. Previously, some efforts were made towards this direction [2], while, the related experimental results have been seriously questioned and debated [3]. Recently, room temperature superconductivity at near-ambient pressure (1 GPa) was claimed in the nitrogen-doped lutetium hydride [4]. Comparing with the previous C-H-S system [2], the pressures required to achieve the claimed room temperature superconductivity in N-doped lutetium hydride is greatly reduced. Actually, superconductivity in hydrogen-rich compounds has been well studied both theoretically and experimentally [5-12]. For example, in $LaH_{10}$ and $YH_9$, researchers have confirmed a credible superconductivity at around 250 K, although a high pressure up to 200 GPa is still needed [13-15]. High-temperature superconductivity of hydrogen-rich compounds at high pressures is understandable [16] since the high-pressure environment can help stabilize chemical structures that would not exist at ambient pressure, and the high Debye temperature and strong electron-phonon coupling constant can allow the occurrence of high temperature superconductivity [17, 18].

The claim of near-ambient superconductivity in N-doped lutetium hydride has been closely scrutinized by the community and doubts immediately arise [19-25]. In order to repeat this discovery, efforts have been made by several groups [19, 21, 22]. So far, no positive and supporting evidence have been found. For example, Li *et al.* [19] reported the high-pressure investigation on lutetium polyhydride and found superconductivity with a maximum $T_c$ of about 71 K under the highest pressure of 218 GPa, which is quite different from the results reported in Ref [4]. Shan *et al.* [21] observed a similar sequence of color change in $LuH_2$ under pressures as reported in Ref [4]. However, no superconducting signal was detected in $LuH_2$ down to 1.5 K under the pressure up to 7.7 GPa. Simultaneously, the metallic N-doped lutetium hydrides with dark-blue color have been successfully synthesized by our group [22], and the results of high-pressure resistance and magnetization measurements show no trace of superconductivity in our samples. Although the near-ambient superconductivity is seriously questioned by the community in N-doped lutetium hydride [4], the pressure induced color change from blue to pink and red seems to be a valid effect. In contrast, in our samples with the similar structure with that of Dasenbrock-Gammon *et al*. [4], we found that the dark-blue color of the $LuH_{2\pm x}N_y$ sample maintained up to 5.2 GPa [22]. This discrepancy inspires us to conduct a further study on the pressure dependent evolution of color and resistance in our $LuH_{2\pm x}N_y$ samples.

In this manuscript, we report the pressure induced color change from dark-blue to pink-violet in our N-doped lutetium hydride $LuH_{2\pm x}N_y$ sample under pressures up to 42 GPa, and the color change is reversible when the pressure is released. Accompanying with the color change from dark-blue to pink-violet, the temperature dependent resistance shows a progressively improved metallic behavior. The value of *RRR* increases sharply in the color change region, and the curves of temperature dependent resistance reveal that the electron-phonon scattering is the dominant one in wide temperature region with a typical Fermi-Liquid behavior in the low temperature region under high pressures above 15.5 GPa. Thus, we have witnessed a Fermi liquid metallic state in $LuH_{2\pm x}N_y$ with a pink-violet color. Superconductivity is still absent in $LuH_{2\pm x}N_y$ at all pressures up to 50.5 GPa in our present study.

**Experimental Details**

Detailed synthesis procedures were described in Ref. [22]. $NH_4Cl$ (Alfa Aesar 99.99%) and $CaH_2$ (Alfa Aesar 98%) were mixed in a molar ratio of 2:8 to generate excessive $H_2$ and $NH_3$. And the initial molecular ratio of Lu and N is 3:1. The $LuH_{2\pm x}N_y$ samples were synthesized under a condition of 300°C for 10 hours with a pressure of 2 GPa using piston-cylinder type high pressure apparatus (LP 1000-540/50, Max Voggenreiter).

The crystal structure was checked by powder x-ray diffraction (XRD) measured on a Bruker *D8* Advanced diffractometer with the Cu $K_{\alpha 1}$ radiation. The Rietveld refinements were conducted with *TOPAS 4.2* software [26]. Scanning electron microscopy (SEM) images and energy dispersive spectrum (EDS) were obtained on a scanning electron microscope (Phenom ProX) with an accelerating voltage of 15 kV. The electrical transport measurements under ambient and high pressure were performed on a Quantum Design physical property measurement system (PPMS). A standard four-probe method was employed to measure the electrical resistivity at ambient pressure. The resistance data under high pressure were obtained by using four-probe van der Pauw method with platinum foil as electrodes [27]. A diamond anvil cell (DAC) with 300 μm culets was used to generate the pressure up to 50.5 GPa. The pressure was measured using the ruby fluorescence method at room temperature [28].

**Results and discussion**

Figure 1(a) shows the schematic crystal structure of $LuH_{2\pm x}N_y$. Lu atoms form the face-centered cubic framework, which dominate the X-ray diffraction patterns of $LuH_{2\pm x}N_y$. $H_1$ and $H_2$ represent two different coordination environments of hydrogen. The atomic position and content of nitrogen are still unknown. When the expected hydrogen locations are fully occupied, the above structure becomes standard $LuH_3$ with the Space Group (SG) of $Fm\bar{3}m$. However, $Fm\bar{3}m$-$LuH_3$ is reported to be only stable under high pressures [12]. Due to the difficulty to identify the light elements by XRD measurement, the actual occupation ratio of hydrogen in N-doped lutetium hydrides remains elusive. Figure 1(b) shows the XRD pattern (black spots) of the $LuH_{2\pm x}N_y$ sample at room temperature and the Rietveld refinement curve (red solid line). We can see that the diffraction peaks can be well fitted by the structure in Fig. 1(a) with the space group of $Fm\bar{3}m$ and lattice parameter $a$ = 5.0322 Å, giving a good agreement with previous reports [4, 22]. There are two tiny peaks which belong to the impurity phase of $Lu_2O_3$, which is encountered quite often in the synthesis of Lu related compounds [4, 21].

We also conducted the composition analysis on the as-grown polycrystals. As shown in Fig. 1(c), the chemical composition of the $LuH_{2\pm x}N_y$ sample is determined by the EDS measurements, and the representative molar ratio of Lu: N is about 15: 1. Note that the composition analysis from EDS instrument is not accurate for light elements, and the elements of C and O should originate from the background signal of the sample holder. It is necessary to point out that in our EDS measurements, we can only detect the existence of nitrogen in some measured areas, which means that there is a clear inhomogeneity of nitrogen concentrations in the LuH$_{2\pm x}$N$_y$ sample. We are aware that LuH$_3$ with hexagonal structure (SG: $P\bar{3}c1$) may be obtained under ambient reaction pressure, but the color is black [29, 30]. When the obtained LuH$_3$ reacts with flowing nitrogen gas, the material transforms into pure LuN with cubic structure. In this sense, it is very interesting to investigate the properties of the cubic LuH$_{2\pm x}$N$_y$ at higher nitrogen level. Actually, the related experiments are already under way.

Figure 1(d) displays the temperature-dependent resistivity of an as-grown LuH$_{2\pm x}$N$_y$ sample at ambient pressure. We can see that the sample shows a good metallic behavior, in good agreement with the previous results [22]. It is clear that there is a roughly linear behavior in the intermediate temperature region, which is expected by the electron-phonon scattering. According to the Bloch-Grüneisen formula [31] based on the electron-phonon scattering, the resistivity should exhibit a feature $\rho(T) = \rho_0 + AT^5$ in the low temperature limit with $\rho_0$ as the residual resistivity; in the intermediate temperature region, one should expect a roughly linear behavior due to the electron-phonon scattering. If there is a strong electron-electron scattering in the system, the resistivity should exhibit a quadratic temperature dependence in low temperature limit [32], namely $\rho(T) = \rho_0 + AT^2$, but in the intermediate and high temperature region, the electron-phonon scattering still prevails and dominates. Thus we fit the resistivity data (from 2 K to 60 K) with the formula $\rho(T) = \rho_0 + AT^\alpha$, the fitting results are $\rho_0$ = 0.25 mΩ · cm, A = 0.37 nΩ · cm K$^{-2.8}$, and α = 2.8. Noting that the exponent α = 2.8 is higher than the typical value anticipated by the Fermi-liquid behavior (α = 2), which reflects electron-phonon scattering and the determination may also be influenced by the quite small *RRR*.

In our previous work, we did not see the change of color in the LuH$_{2\pm x}$N$_y$ sample under pressures up to 5.2 GPa, and the dark-blue color is still maintained. Such phenomenon is quite different from what observed by Dasenbrock-Gammon *et al.* [4] in N-doped lutetium hydride (LuH$_{3-\delta}$N$_\varepsilon$), and also in LuH$_2$ reported by Shan *et al.* [21]. In those works, the pressure induced color change from dark-blue to pink and red occurs at a relatively low pressure (1-2.2 GPa). In order to explore whether our LuH$_{2\pm x}$N$_y$ sample has color change under higher pressures, we used an optical microscope to directly detect the color of the samples under different pressures through a DAC. Surprisingly, we also observed a continuous color transformation from dark-blue to pink-violet of our LuH$_{2\pm x}$N$_y$ sample starting from about 12 GPa, as shown in Fig. 2. And it completely changes to pink-violet color at about 21 GPa. With further increasing the pressure, the sample remains pink-violet color up to 42 GPa, the maximum pressure in this measurement for the color image. Moreover, after releasing the high pressure, the color returns to the original dark-blue one. Thus, the color change in LuH$_{2\pm x}$N$_y$ sample is also reversible. Except for the higher pressures required for the color change, the observations are quite similar to the N-doped Lu hydride (LuH$_{3-\delta}$N$_\varepsilon$) [4] and LuH$_2$ [21]. Thus we speculate that there is a structural transition induced by pressure. The exact reason for requiring the higher pressures for this color change may be attributed to the different content of N in LuH$_{2\pm x}$N$_y$ samples.

To investigate whether the color change has a significant effect on the transport property, in Fig. 3(a), we show the temperature dependent electrical resistance of LuH$_{2\pm x}$N$_y$ in the pressure range of 9 ~ 50.5 GPa. Similar to the situation at low pressures, a hump structure around 300 K and a weak upturn in the low temperature region in $R(T)$ curves are observed at 9.0 and 12.6 GPa. Under higher pressures, such features are gradually suppressed. Besides, accompanied with the decrease of the absolute value of electrical resistance, the system transforms from a bad metal (inferred from the slight upturning of resistivity at low temperatures and small *RRR*) to an optimized metallic behavior under higher pressures. This tendency could be more visualized through the change of $R/R_{300K}$ as shown in Fig. 3(b), in which the decrease of resistance in the whole temperature region becomes more and more explicit with increasing pressure. To our surprise, when the color of the sample has been completely transformed to pink-violet, the $R(T)$ curves of the sample even shows a better metallic behavior. This is completely different from our usual intuitions, namely, a material with dark color should exhibit a better electric conductivity. This, of course, remains to be further explored. To figure out whether there is a possible magnetoresistance (MR) in the sample with pink-violet color, we present the magnetic field dependence of the resistance under 50.5 GPa at 10 K and 100 K in Fig. 3(c). The results reveal a negligible MR effect up to 9 T. This suggests that the electric conduction is dominated by a single band, since in a multiband system, the MR should be quite significant [33]. As evidenced by the data in Fig.3(a), we still do not observe a superconducting related transition or zero-resistance state of $LuH_{2\pm x}N_y$ in our electrical resistance measurements down to 2 K with pressures up to 50.5 GPa.

Under pressures above about 15.5 GPa, the resistivity shows a power-law behavior in low temperature region, a roughly linear behavior in the intermediate temperature region, and a slightly bending down in high temperature region. All these are consistent with the expectation of electron-phonon scattering. In order to understand the electronic properties of the system under different pressures, we made a fitting to $R(T)$ curves above 15.5 GPa at low temperatures. The red solid lines in Fig. 3(a) represent a fit based on the equation $R(T) = R_0 + AT^\alpha$, where $R_0$ represents the residual resistance, the coefficient $A$ and the exponent $\alpha$ are associated with the density of states at Fermi level and the inelastic electron scattering rate, respectively [34]. However, the above equation is not suitable to describe the resistance data at 9.0 and 12.6 GPa because of the upturn of resistivity at low temperatures, which may be related to strong scattering by the grain boundaries of the polycrystalline sample [35].

We summarized our results in Fig. 4. At low pressures (below 10 GPa), the resistance at 200 K, 100 K and 10 K almost overlaps with each other and decreases gradually towards higher pressure. After that, all of them drop abruptly in the pressure range of 12 ~ 20 GPa and finally become almost flattened (see Fig. 4(a)). On the other hand, the pressure dependent *RRR* shows an opposite tendency, where a sudden enhancement of *RRR* was observed in the pressure range from 12 GPa to 20 GPa. We need to point out that this pressure range is exactly in the region where the color change happens.

Moreover, by extracting the fitting parameters in $R(T) = R_0 + AT^\alpha$, we plot pressure dependent $R_0$ and the exponent α in Fig. 4(b). With the increase of pressure, the residual resistance $R_0$ decreases monotonically. And the value of $\alpha$ decreases from 2.1 to 1.8. indicating that the transport behavior of $LuH_{2\pm x}N_y$ at low temperatures is roughly consistent with Fermi liquid behavior with the pressure above 15.5 GPa, although the electron-phonon scattering is still dominant in the intermediate and high temperature regions. By carefully examining the data plotted in Fig. 4(b), we also found a clear change in the slope of the curves with the pressure about 20 GPa, which corresponds exactly to the pressure at which the material color changes to pink-violet. All these may indicate that an interesting transition occurs with the color change of the $LuH_{2\pm x}N_y$, which will stimulate further studies.

**Conclusion**

In summary, we have observed the color change of the $LuH_{2\pm x}N_y$ sample from dark-blue to pink-violet under high pressure, and the transformation occurs over a wide pressure range from 12 GPa to 21 GPa. Accompanied with the change of material color, the transport behavior of $LuH_{2\pm x}N_y$ sample also changes significantly, including the greatly optimized metallic behavior, clear enhancement of *RRR,* reduced residual resistance $R_0$. Although the resistivity seems to be dominated by the electron-phonon scattering in the intermediate and high temperature regions, the low temperature data indicate the appearance of Fermi-liquid behavior. All these suggest that the $LuH_{2\pm x}N_y$ system may undergo a possible transition in this pressure region. The metallic state at high pressures shows a pink-violet color with a Fermi liquid feature. Our resistance measurements of the $LuH_{2\pm x}N_y$ sample up to 50.5 GPa reveal the absence of superconductivity down to 2 K.

**Acknowledgements**

This work was supported by the National Key R&D Program of China (No. 2022YFA1403201), National Natural Science Foundation of China (Nos. 12061131001, 12204231, 52072170, and 11927809), and Strategic Priority Research Program (B) of Chinese Academy of Sciences (No. XDB25000000).

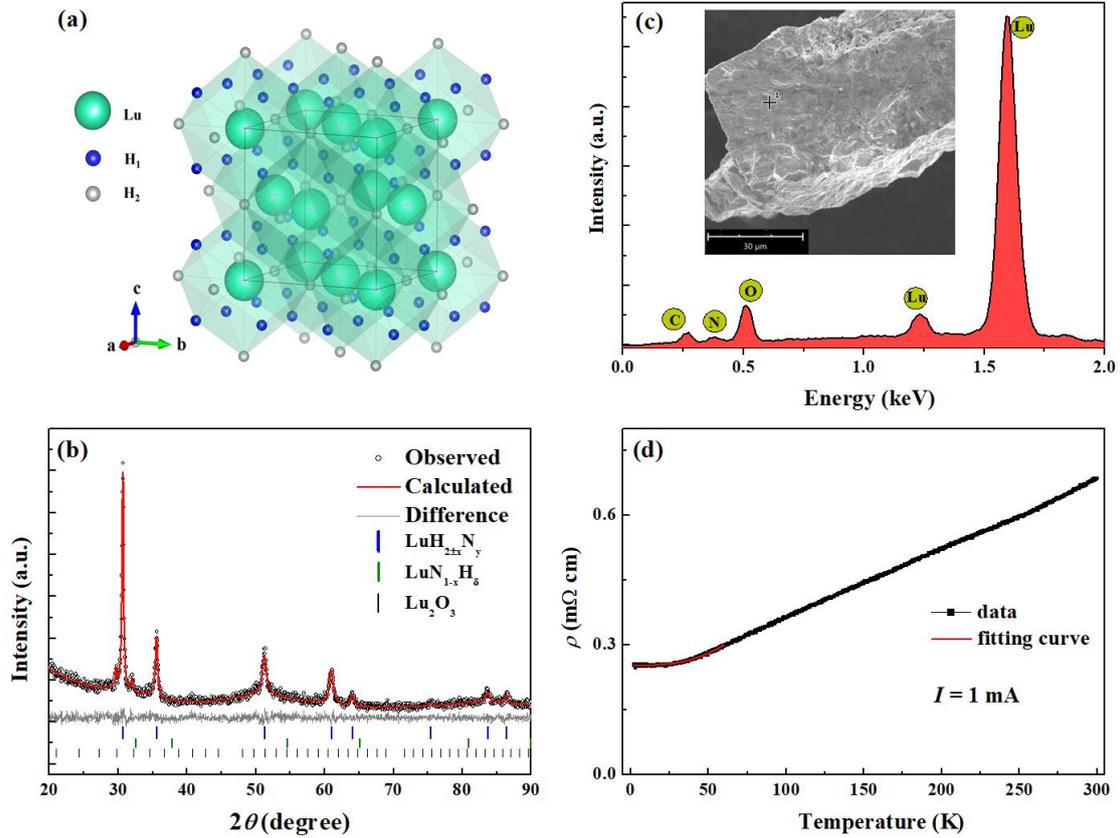

**Figure 1.** (a) Schematic crystal structure of $LuH_{2\pm x}N_y$. Lu, $H_1$, $H_2$ atoms are shown by green, blue and grey spheres, respectively. The different colors of H atom represent the two possible coordination environments. (b) Powder x-ray diffraction patterns and Rietveld fitting curve of $LuH_{2\pm x}N_y$. The vertical bars are the calculated Bragg reflection positions for $LuH_{2\pm x}N_y$, $LuN_{1-x}H_\delta$ and $Lu_2O_3$. (c) The typical energy dispersive spectroscopy (EDS) of $LuH_{2\pm x}N_y$. Inset shows the SEM image of the polycrystalline sample in the secondary electron mode. (d) Temperature dependence of the electrical resistivity of $LuH_{2\pm x}N_y$ measured in the temperature range from 2 K to 300 K at ambient pressure. The red solid line is a fitting curve by using the equation $\rho(T) = \rho_0 + AT^\alpha$.

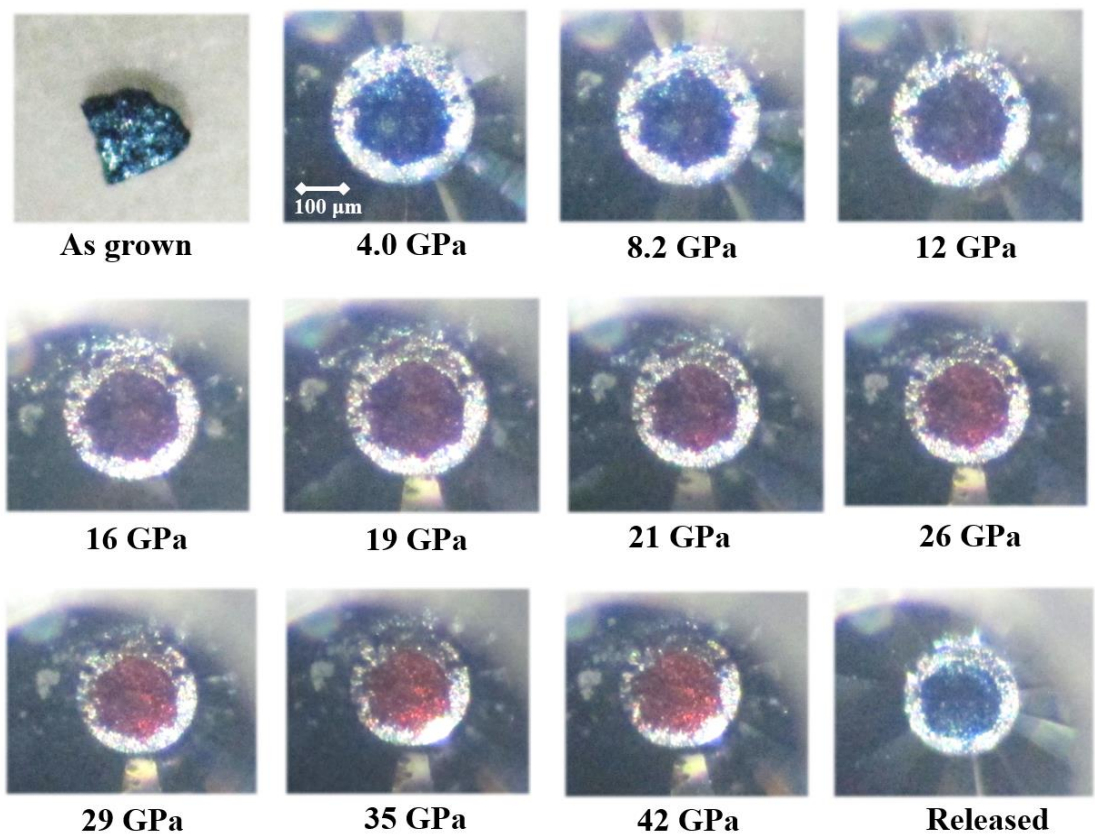

**Figure 2.** The optical microscope images of LuH$_{2\pm x}$N$_y$ in a DAC chamber at different pressures up to 42 GPa. Dark-blue color begins to change at about 12 GPa and it completely changes to pink-violet color at about 21 GPa.

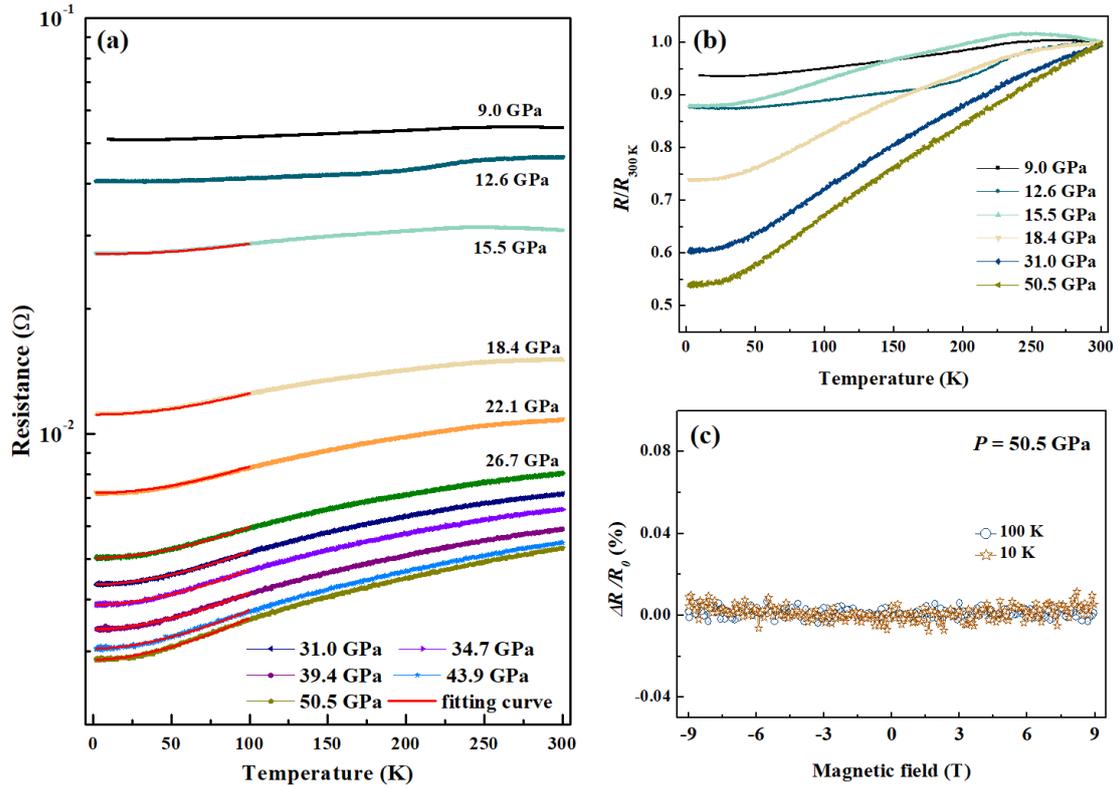

**Figure 3.** (a) Electrical resistance of LuH$_{2\pm x}$N$_y$ under various pressures up to 50.5 GPa. The red solid lines represent the fitting curves using the equation $R = R_0 + AT^\alpha$ in the temperature range from 2 K to 100 K. (b) Normalized $R-T$ curves of LuH$_{2\pm x}$N$_y$ at selected pressures from 2 K to 300 K. (c) Field dependence of magnetoresistance of LuH$_{2\pm x}$N$_y$ measured at 10 K and 100 K under the pressure of 50.5 GPa, where $\Delta R$ is $R - R_0$, and $R_0$ is the resistance at zero field.

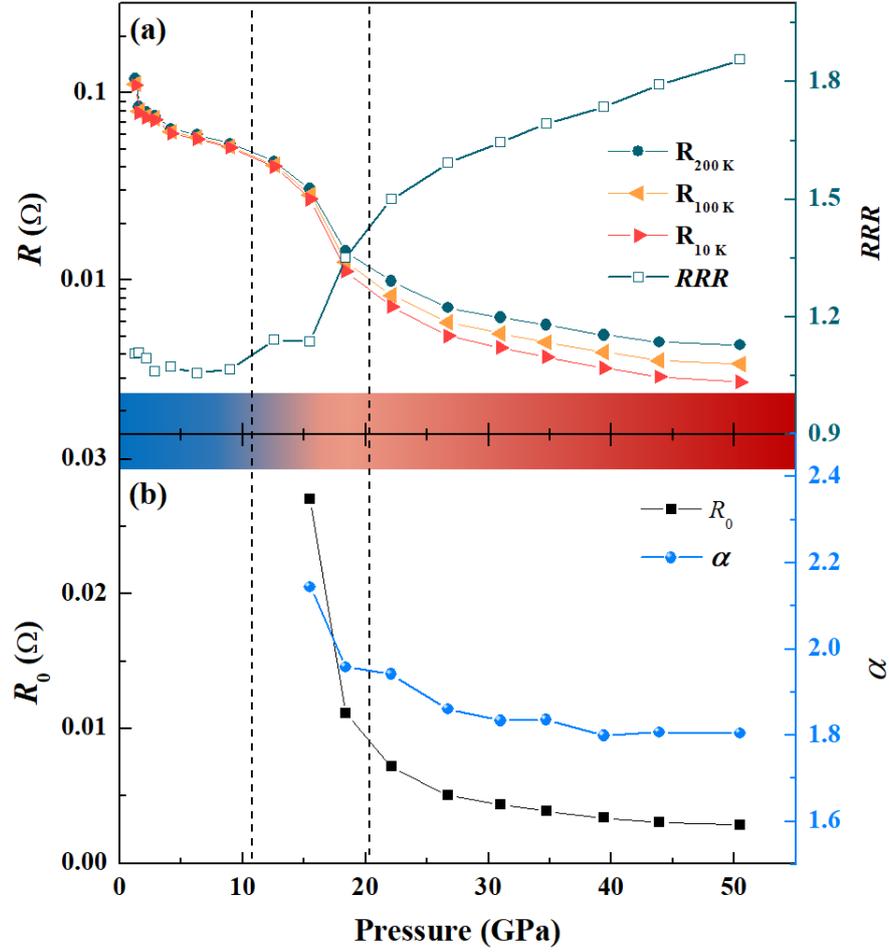

**Figure 4.** (a) Pressure dependence of resistance at 200, 100m and 10 K, namely $R_{200K}$, $R_{100K}$, $R_{10K}$, and $RRR$. (b) Pressure dependence of $R_0$, the exponent $\alpha$ derived from the fitting to $R(T) = R_0 + AT^{\alpha}$ of LuH$_{2\pm x}$N$_y$ in low temperature region. The middle of the figure shows a schematic mapping of color changing with pressure and the two vertical dashed lines show the pressure region in which the color change occurs. The data below 9 GPa in (a) are extracted from Ref. [22].